\documentclass[aps,prl,twocolumn,showpacs,superscriptaddress,groupedaddress]{revtex4} 

\usepackage{graphicx}
\usepackage{subfigure}

\usepackage{amsmath} 

\usepackage[colorlinks=true,citecolor=blue,urlcolor=magenta,breaklinks]{hyperref}
\usepackage{natbib}
\usepackage[T1]{fontenc}
\usepackage{times}
\usepackage{tikz,xcolor,hyperref}
\definecolor{lime}{HTML}{A6CE39}
\DeclareRobustCommand{\orcidicon}{
	\begin{tikzpicture}
		\draw[lime, fill=lime] (0,0) 
		circle [radius=0.16] 
		node[white] {{\fontfamily{qag}\selectfont \tiny ID}};
		\draw[white, fill=white] (-0.0625,0.095) 
		circle [radius=0.007];
	\end{tikzpicture}
	\hspace{-2mm}
}
\foreach \x in {A, ..., Z}{%
\expandafter\xdef\csname orcid\x\endcsname{\noexpand\href{https://orcid.org/\csname orcidauthor\x\endcsname}{\noexpand\orcidicon}}
}

\begin{document}


%
\title{Axion spectra and the associated x-ray  spectra of low-mass stars}
\author{Il\'idio Lopes\orcidA}
\email[]{ilidio.lopes@tecnico.ulisboa.pt}
\affiliation{Centro de Astrof\'{\i}sica e Gravita\c c\~ao - CENTRA,
Departamento de F\'{\i}sica, Instituto Superior T\'ecnico - IST,\\ 
Universidade de Lisboa - UL, Avenida Rovisco Pais 1, 1049-001 Lisboa, Portugal \\}

%
%
%
\begin{abstract}
Axion particles are among the best candidates to explain dark matter and resolve the strong $CP$ problem in the Standard Model. If such a particle exists, the core of stars will produce them in large amounts.  
For the first time, we predict the axion spectra and their associated luminosities for several low-mass stars --  one and two solar masses stars in the main sequence and post-main sequence stages of evolution. Equally, we also compute the x-ray excess emission resulting from the conversion of axions back to photons in the presence of a strong magnetic field in the envelope of these stars.
Hence, a given star will have a unique axion spectrum and $L_a$ axion luminosity. And if such star has a strong magnetic field in its stellar envelope, it will also show a characteristic x-ray spectrum and $L_{a\gamma}$ x-ray luminosity.  Such radiation will add up to the x-ray electromagnetic spectrum and 
$L_X$
luminosity of the star. The present study focuses on axion models with an axion-photon coupling constant, $5\;10^{-11} {\rm GeV}^{-1}$, a value just below the most recent upper limit of $6.6\;10^{-11} {\rm GeV}^{-1}$  found by CAST and IAXO helioscopes.
The range of axion parameters discussed here spans many axion models' parameter space, including the DFSZ and KSVZ  models.
We found that axions with a mass in the range $10^{-7}$  to $10^{-5}\; {\rm eV}$ and an axion-photon coupling constant of $5\;10^{-11} {\rm GeV}^{-1}$  produce an axion emission spectra with an averaged axion energy that varies from 1 to 5 KeV, and an  $L_{a}$  ranging from $10^{-6} $ to $7\;10^{-3} \;L_\odot$.  
We also predict that $L_{a\gamma}$ varies from $5\;10^{-8}$ to $10^{-5} \;L_\odot$ for stars with   an averaged magnetic field of $3 \;10^{4}\;{\rm G}$ in their atmospheres. 
Most of these $L_{a\gamma}$  predictions are larger than the $L_X$   observed in some stars. Therefore, such preliminary results show the potential of the next generation of stellar x-ray missions to constrain several classes of axion models.
\end{abstract}


\maketitle


\section{I. Introduction\label{sec-intro}}

\medskip\noindent
The axion or their closest relatives -- the axionlike particles (ALP) are among the most popular candidates that are being proposed to explain the existence of dark matter.  An ALP is a ultralight pseudoscalar boson $a$ predicted by several extensions of the Standard Model
\citep[SM, e.g.,][]{2020PhR...870....1D}. ALPs are well-motivated particles that occur in many extra-dimension theories, like string theory and notably the M theory \citep[e.g.,][]{1996PhRvL..76.1015T,2016PhR...643....1M}.  Nowadays, all these types of particles capable of solving the strong $CP$ problem are many times referred to as QCD axion \citep{2019PhRvD..99c5037D}. In this article, we opt to call all these different kinds of particles, ALPs included, simply  by "axions"  if not stated otherwise.

\medskip\noindent
Many of these SM extensions predict the existence of axion channels' with photons $\gamma$ and electrons $e$ \citep{1977PhRvL..38.1440P,1978PhRvL..40..279W,1978PhRvL..40..223W}; such processes are regulated by the coupling constants $g_{a\gamma}$ or $g_{ae}$ respectively. Two large classes of axion models are currently quite popular in the literature,      the Dine-Fischler-Srednicki-Zhitnitsky \citep[DFSZ, e.g.,][]{1981PhLB..104..199D}  and the Kim–Shifman–Vainshtein–Zakharov \citep[KSVZ, e.g.,][]{1979PhRvL..43..103K,1980NuPhB.166..493S}  in which the anomaly is carried by SM quarks or new colored fermions, respectively \citep[e.g.,][]{2010AIPC.1200...83K}.  
The two axion channels mentioned above occur in DFSZ and KSVZ models, although the axion-electron interaction is more common in the grand unification theories or GUTs
 \citep[e.g.,][]{2018PrPNP.102...89I}.
In astrophysical scenarios both processes can exist, the dominant spectrum will depend mostly on the magnitude of the relative values of $g_{ae}$ and $g_{a\gamma}$ \citep[e.g.,][]{2020PhR...870....1D}.  
In this work we will focus on the latter process.

\medskip\noindent
The axion-photon channel gives rise to the Primakoff production of axions inside stars. Indeed, \citet{1951PhRv...81..899P} discovered that in the presence of external electric and magnetic fields photons convert into axions. Although many works compute the axion emission spectrum for the Sun, this has not yet been done for other low-mass stars. 
A detailed calculation of the solar axion emission spectrum is found in \citet{2007JCAP...04..010A} and references therein. A more   detailed study by  \citet{2020PhRvD.102l3024G}  found that solar large-scale magnetic fields  enhance the axion emissivity by the coherent conversion of thermal photons to axions.  It was also reported that, in the Sun, the plasmon-axion conversion could compete with the Primakoff production of axions \citep{2020PhRvD.102d3019O}. Equally, \citet{2020PhRvD.101l3004C} found that, depending on the strength and nature of the solar magnetic field, the plasmon-axion conversion dominates over Primakoff production for energies lower than $200 \; eV$.

\medskip\noindent
Here, we are interested in studying for the first time the axion spectrum emitted inside low-mass stars and the electromagnetic spectrum excess produced by the axion-to-photon conversion in the atmosphere of the same stars.  Accordingly, we will compute the Primakoff production of axions inside stars and their transformation back to x-ray radiation by inverse Primakoff production in the stellar surface. 
Since we compute the spectra of axions in the energy range from 1 to 15 KeV, the contribution of addition mechanisms mentioned above and in the Refs.
\citep{2020PhRvD.102l3024G,2020PhRvD.102d3019O,2020PhRvD.101l3004C}
is negligible.
This study is of interest for the next generation of stellar x-ray observational missions, for which the high precision measurement of the electromagnetic field will make it possible to put a stringent constraint on several classes of axion models. Low-mass stars with masses varying from 1 to 2 solar masses are good observational targets 
to study the impact of axions on the stellar evolutions. Among other reasons, we highlight the following ones:  (i) these stars are very abundant in the Universe;
(ii) the internal structure of these stars, including our own Sun, is very well known (by means of helio- and asteroseismology), much better than for more massive stars (which are much less abundant in the Universe); (iii) the atmosphere of these stars and their magnetic fields are well understood.

 \medskip\noindent
We organize the article in seven sections: In Sec. II we explain the motivation for the axion existence in particle physics, cosmology and astrophysics; in Sec. III, we motivated the interest of using low mass stars to study the axion properties; in Sec. IV we explain how the axions are produced inside  low-mass stars, and compute the axion spectrum for these stars in the  main-sequence and post-main sequence phases; in Sec. V we present the model used to calculate the x-ray spectrum of axions; in Sec. VI we analyse and discuss  the x-ray emission from axions for several stellar models. And finally, in Sec. VII, we summarize and present our results.
If not stated otherwise, we work in natural units where the speed of light in the vacuum, the Planck constant and the Boltzmann constant are set to unity  $\hbar=c=k_B=1$.

\section{II. Axion motivation in  particle physics and astrophysics}

\medskip\noindent
The axion appeared originally on a pioneer generalization of the standard model that was made by introducing a new global symmetry known now as the Peccei-Quinn symmetry \citep{1977PhRvL..38.1440P}. Such transformation allows to explain the absence of the $CP$ violation in the strong interaction (the so-called QCD strong $CP$ problem).  
This first model was proposed by \citet{1977PhRvL..38.1440P}, as well as by
\citet{1978PhRvL..40..223W}  and \citet{1978PhRvL..40..279W} and since then is known as the Weinberg-Wilczek-Peccei-Quinn model. The axions in this model have a symmetry-breaking scale of the order of the electroweak scale. Since this model has been ruled out, many alternative models (usually referred to as ALP models)  are being developed where the symmetry-breaking scale becomes arbitrary.

\medskip\noindent
The success of the axion theories has motivated many experimental physicists to propose and develop experiments to search for such particles: several detectors around the world are actively searching for the axions
using the inverse Primakoff effect. The most well known of these detectors is the CERN Axion Solar  Telescope or CAST \citep[i.e.][]{2001PhLB..515....6B}.
The most recent upper limit obtained  CAST  fixed the upper limit of the axion-photon coupling in $6.6\;10^{-11} {\rm GeV}^{-1}$
at 95\% confidence level \citep{2017NatPh..13..584A}.
Moreover, a new generation of experiments is being built to continue the CAST research, like International Axion Observatory \citep[i.e., IAXO,][]{2014JInst...9.5002A} or  the dielectric haloscope MADMAX \citep[i.e., MAgnetized Disk and Mirror Axion eXperiment  ,][]{2020EPJC...80..392E,2018JCAP...11..051K}. 
However many other detectors based on  different experimental techniques and strategies have been proposed during the last decade to look for axions, like multilayer optical haloscopes \citep{2018PhRvD..98c5006B}, plasma haloscopes \citep{2019PhRvL.123n1802L}, dish antennae \citep{2013JCAP...04..016H},
and  a new class of detectors based on 
Antiferromagnetically doped topological insulators
\citep[TOORAD;][]{2019PhRvL.123l1601M}.

\medskip\noindent
Axions are excellent candidates to explain many features that occur during the formation and evolution of the Universe, including inflation, dark radiation and  dark energy \citep[e.g.,][]{2018JCAP...10..005W}. The production and existence of axions has been suggested in many cosmological and astrophysical contexts, such as  pure axion stars and hybrid axion stars ~\citep[i.e.,][]{2018PhRvD..98b3009C,2018PhLB..777...64V,2019IJMPD..2850111P}, to mention a few. 
In some scenarios, the gravitational field of a recently formed axion star can capture neutral hydrogen leading to the appearance of a hypothetical mix of stars ~\citep{2016JHEP...12..127B}. Unlike pure axion stars, due to the electromagnetic emission these should be easy to observe. 
Axions are now known to produce many other astrophysical phenomena. 
For instance, they produce an unique electromagnetic spectrum stimulated by the presence of strong magnetic fields during the core-collapse of a supernova \citep{2015JCAP...02..006P}, or trigger superradiant instability of spinning black holes  \citep{2018JCAP...03..043C}.

\section{III. Axions, stars and magnetic fields}
\label{sec-ASMF}

\medskip\noindent
The impact of axions on stars has already been discussed for a diverse population of stellar objects. In general, the presence of axions inside stars leads to the formation of efficient energy-loss mechanisms. Due to this it has been possible to put constraints to the properties of such particles, for instance using the Sun \citep{1989PhRvD..39.2089V}, red giant stars\citep{2013PhRvL.111w1301V}, white dwarfs \citep{2019PhRvL.123f1104D} and neutron stars \citep{2012ApJ...748..116P}. There are already many observations of stars at different stages of their evolution that show trends of high energy losses and unique features in their electromagnetic spectra. In many of these cases, the standard cooling mechanisms are unable to explain such astronomical phenomena. The only way to describe these observations is to propose non-standard cooling channels; for which axion emission is a very plausible solution \citep[e.g.,][]{1992ApJ...392L..23I,2011PhRvD..84j3008R,2016JCAP...05..057G,2017JCAP...10..010G}.

\medskip\noindent
The fact that the structure of  low-mass stars is much better known 
than other stars also results from the high-quality data already
available and possible to be obtained by the current \citep{2008Sci...322..558M,2010ApJ...713L..79K} and future
\citep{2014ExA....38..249R}  asteroseismology missions. 
Asteroseismology probes the internal structure of these stars with unprecedented precision. 
These high-quality observations allow to obtain a detailed knowledge of the thermodynamical structure of many low-mass stars at different stages of evolution: main-sequence, sub-giant and red-giant phases. Such a unique set of asteroseismic data combined with high-precision spectroscopic observations make such stars a unique precision laboratory of experimental physics study to the properties of new particles like axions.

\medskip\noindent
Particularly relevant for this study is the existence of magnetic fields in these stars. In the last seven decades powerful techniques based on the Zeeman effect \citep[e.g.,][]{1947ApJ...105..105B,2009ARA&A..47..333D}
and spectropolarimetry \citep[e.g.,][]{1997MNRAS.291..658D}
have made possible the detection and measurement of magnetic activity of several star's types across the HR diagram. Recent measurements found that main sequence and sub-giants stars with masses below 2.5 $M_\odot $ have surface magnetic fields that vary between $0.5$ and $40 \; {\rm kG}$ \citep[e.g.,][]{2009IAUS..259..323B,2019A&A...621A..47K,2019arXiv191207241K}. Classical examples of  magnetic main-sequence stars are HD 215441 ($B\sim 3.4 \times 10^4\;{\rm G}$), 
HD 154708 ($B\sim 2.45 \times 10^4\;{\rm G}$), 
HD 137509 ($B\sim 2.9 \times 10^4\;{\rm G}$), 
and HD 75049 ($B\sim 3 \times 10^4\;{\rm G}$)
\citep[see][and references therein]{2015SSRv..191...77F}.
Such stellar activity usually is accompanied  by a significant x-ray emission, i.e.,  a large luminosity in the x-ray energy band of the electromagnetic spectrum. 
\citet[][]{2016MNRAS.462.4442S} found that  for main-sequence stars  with masses smaller than 1.5 $ M_\odot$, $L_{\small X}$ varies from $10^{-6}$ to $10^{-4}\;L_\odot$, where emissions have a sinusoidal  behavior  with periods varying from a few days to a month. This intense x-ray radiation is particularly prominent 
on the chemically peculiar $A_p$ and $B_p$ stars \citep{2016AdSpR..58..727R}. Since dynamo processes are at the origin of magnetic fields in the outer layers of all these stars  \citep[e.g.,][]{2014SSRv..186..535L}, for which their intensities change from the equator to the poles of the star and over long stellar cycles \citep{2017AN....338...26N,2019Ap.....62..177S}.
\citet{1995ApJ...438..269B} reported for the first time in a systematic manner the existence of sun-like magnetic cycles on low mass stars by measuring the chromospheric variations on these stars.
 Consequently, most of the x-ray emission in these stars will be strongly time-dependent. For some of these stars it is even possible  to observe such
x-ray time variability.  Recently the XMM-Newton satellite has discovered such a long-term X-ray magnetic cycle in $\epsilon$-Eri \citep{2020A&A...636A..49C}.
Such x-ray time dependence results from the variability of the stellar magnetic field. As we will discuss later,  such magnetic field time dependence induces a variation on the x-ray spectrum resulting from axion conversion to photons in the stellar atmosphere.

\section{IV. Axion Primakoff emission spectrum in stars}

\begin{figure*}
	\centering
	\begin{tabular}{cc}
		\includegraphics[width=70mm]{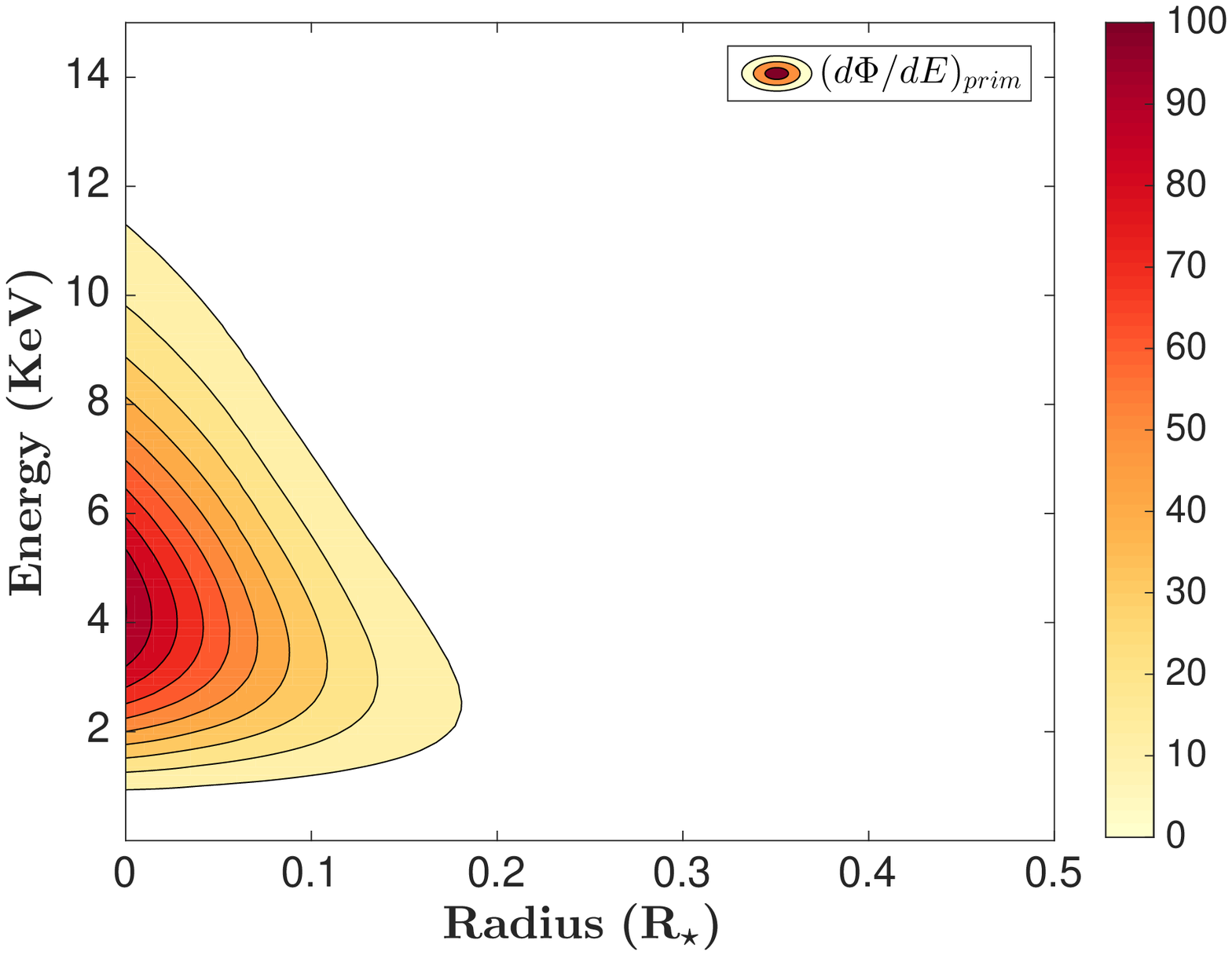}
		& \hspace{-1.0cm}	
		\includegraphics[width=70mm]{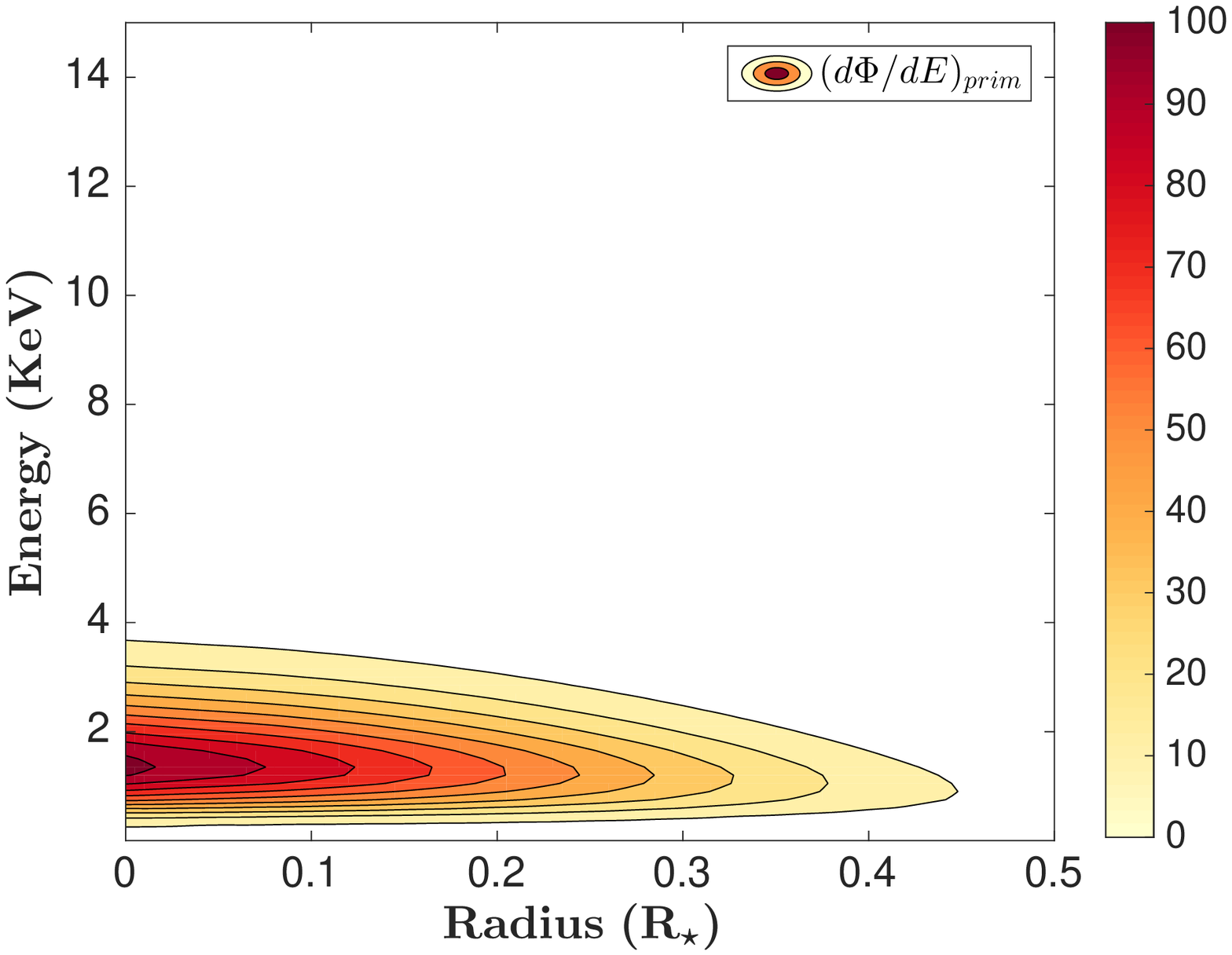}
		\\
		\\
		\vspace{-0.5cm}	
		\\
		\includegraphics[width=70mm]{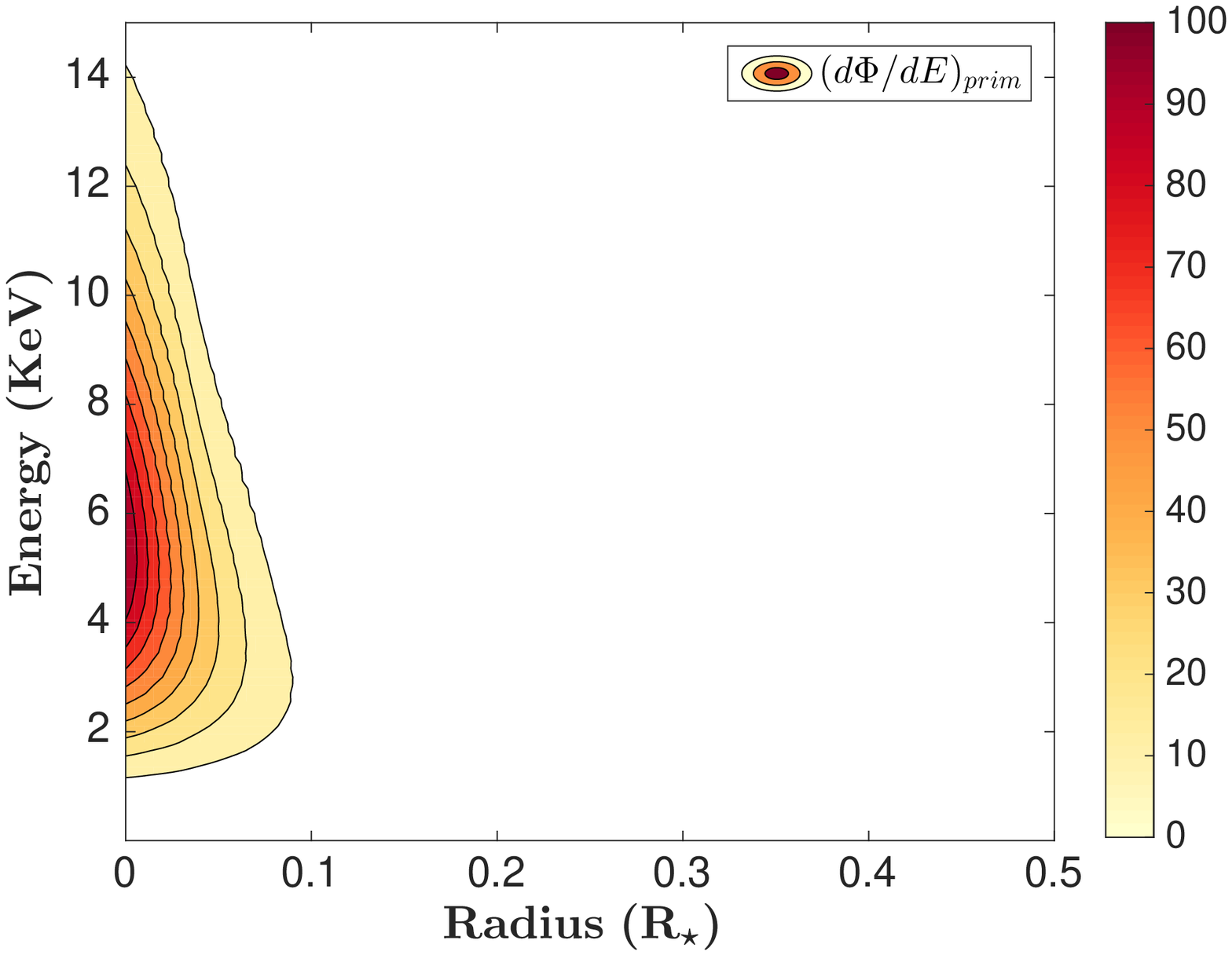}
		&\hspace{-1.0cm}	
		\includegraphics[width=70mm]{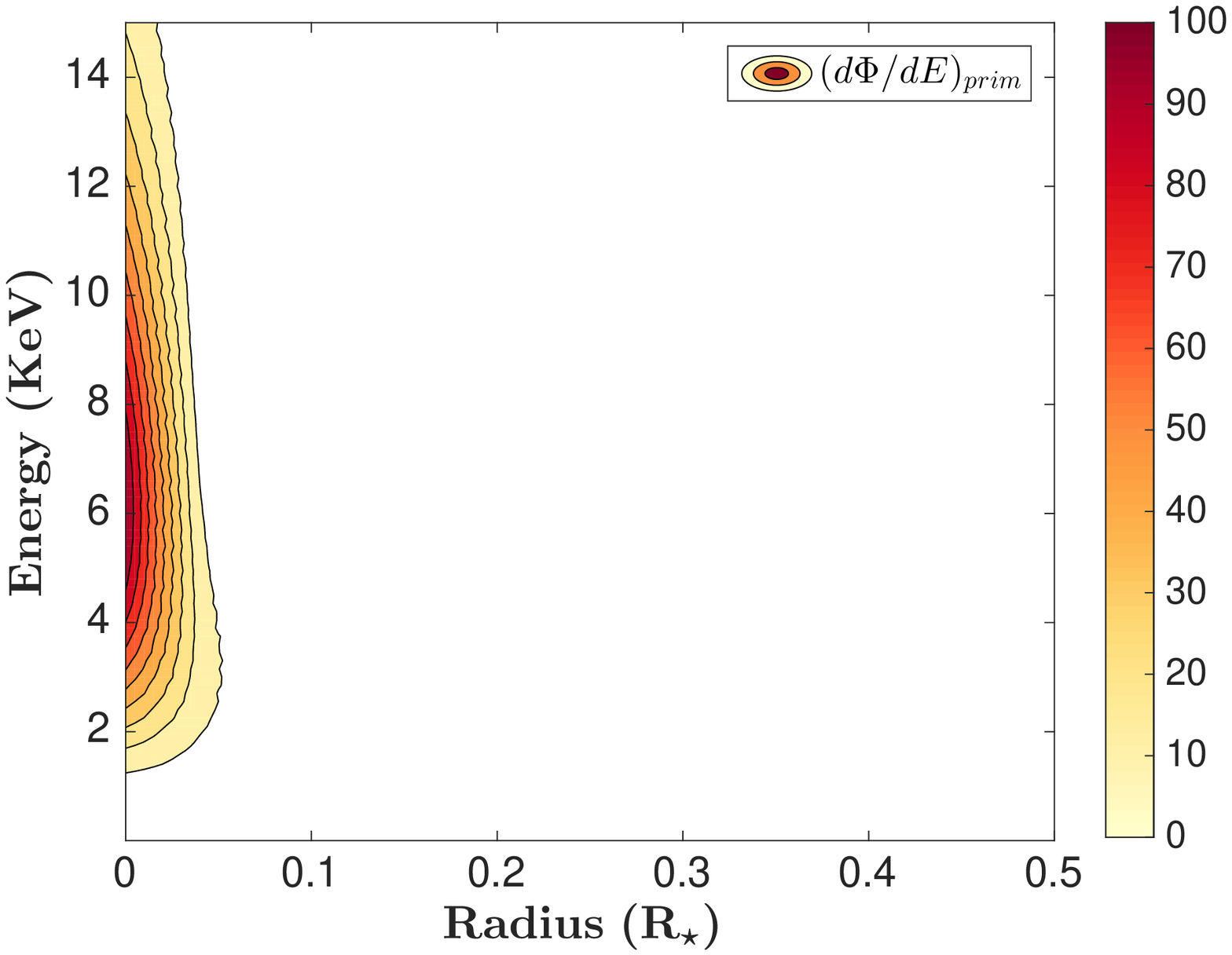}\\
	\end{tabular}
	\vspace{0.2cm}
	\caption[]{The radial distribution of the axion energy loss rate $\mathbf{(d\Phi/dE)_a}$  
		in the core of the star for the models: {\bf top panels}  - $A_\odot$ (Sun) and $A_1$; 
		{\bf lower panels} - $A_2$  and $B_3$.  The description of the models is on Table \ref{tab:axionstars}. The contours correspond to the variation of $\mathbf{(d\Phi/dE)_a}$  (as a percentage) between the centre and the surface of the star. In this representation, $\mathbf{(d\Phi/dE)_a}$ is independent of $g_{a\gamma}$.}
	\vspace{1.2cm}
	\label{fig:Primakoff_inside_r}
\end{figure*}

The phenomenology commonly used to describe all axions (including ALPs) result from these being very light particles with relatively low masses and very weak interactions. These particles interact so weakly with other particles that they
have in the past been referred to as invisible \citep[e.g.][]{1979PhRvL..43..103K,1981PhLB..104..199D}.
Axions behave similarly to neutrinos, like them they provide an additional  cooling mechanism in stars, their impact on the evolution of a star is well documented \citep[e.g.,][]{2008LNP...741...51R,2015JCAP...05..050A}.

\medskip\noindent
The interior of stars is a relatively well understood weakly coupled plasma which permits the precise calculations of axion production reactions \citep{2010AIPC.1200...83K}. Although many different types of axion reactions occur inside stars,  the most crucial axion reaction is the axion-to-photon coupling --  Primakoff  reaction, that drives the  production of axions in photon collisions with charged particles $(Z)$ in the stellar plasma: $\gamma+Z\rightarrow a+Z$ \citep[e.g.,][]{1988PhRvD..37.1356R,2018PrPNP.102...89I}. 
This process that produces (or annihilates) axions reads
\begin{eqnarray}
{\cal L}_{a\gamma}=-\frac{g_{a\gamma}}{4}F_{\mu\nu}F^{\mu\nu}a
=g_{a\gamma} \mathbf{E}\cdot  \mathbf{B}\, a,
\end{eqnarray}
where $a$ is the axion field, $F$  and $\tilde{F}$ the electromagnetic field-strength tensor and its dual, $\mathbf{E}$ and $\mathbf{B}$ are the electric and magnetic fields, and $g_{a\gamma}$ the axion-photon coupling constant. Such interaction implies the conversion of $\gamma$ into $a$ (and its reverse) in the presence of external electric and magnetic fields.

\medskip\noindent
The hot plasma of the interior of most  stars is an excellent source for the production of axions with an energy of several keV, by the transformation of thermal x-ray photons into axions in the electric fields of the charged particles \citep{1988PhRvD..37.1356R}.
The transition rate for a photon of energy $E$ into an axion of the same energy by the Primakoff effect in a stellar plasma is 
\begin{eqnarray}
\Gamma_{\gamma a}=
\frac{g_{a\gamma}^2 T k_s^2}{32\pi}
\left[\left(1+\frac{k_s^2}{4E^2}\right)
\ln{\left(1+\frac{4E^2}{k_s^2}\right)}-1\right],
\label{eq:Gamma_axion}
\end{eqnarray}
where T is the temperature and $k_s$ the screening scale.  
$k_s$
 in the Debye-H\"uckel approximation is
\begin{eqnarray}
k^2_s=\frac{4\pi \alpha_{em}}{T}
\left(n_e+\sum_{nuclei}Z_j^2n_j\right),
\label{eq:k_axion}
\end{eqnarray}
where $\alpha_{em}$ is the fine-structure constant,
 $n_e$ and $n_j$ are the number densities of electron 
and of the jth ion of charge $Z_j$.  Thus, the axion flux spectrum produced by the star up to a certain dimensionless   radius  ($0 \le r\le 1$)  is obtained from
\begin{eqnarray}
\left(\frac{d\Phi}{dE}\right)_a=
2\pi\int_{0}^r \varphi_a (r,E) \; r dr,
\label{eq:dPhidE}
\end{eqnarray}
where $r=R/R_\star$ and $R_\star$ is the total stellar radius.
$\varphi_a (r,E)$ is the stellar axion luminosity at the 
dimensionless   radius  \citep{2007JCAP...04..010A}, accordingly
\begin{eqnarray}
\varphi_a (r,E)=\frac{R_\star^3}{2\pi^3D_{\rm AU}^2}
\int_r^1
\frac{E k \Gamma_{\gamma a}}{e^{E/T}-1}
\frac{xdx}{\sqrt{x^2-r^2}},
\label{eq:varphi_a}
\end{eqnarray}
where  $D_{\rm AU}$ is the fiducial distance from the Earth to the star. The wave number $k$ is
\begin{eqnarray}
k^2=E^2-\omega_{pl}^2,
\label{eq:Kappa}
\end{eqnarray}
where  $\omega_{pl}$ is
 the electron plasma frequency. This last quantity is given by
 \begin{eqnarray}
  \omega_{pl}=\sqrt{\frac{4\pi\alpha_{em}n_e}{m_e}},
  \label{eq:omegapl}
\end{eqnarray}
 where $m_e$ is the electron mass. The total axion flux at the 1 A.U. is
\begin{eqnarray}
\Phi_a=2\pi \int_0^1 \int^\infty_{\omega_{\rm pl}} \varphi_a (r,E)\, dE\,   r dr,
\label{eq-Phi_a}
\end{eqnarray}
the mean axion energy is 
\begin{eqnarray}
\langle E_a\rangle=\frac{2\pi}{\Phi_a}
\int_0^1 \int^\infty_{\omega_{\rm pl}} \varphi_a (r,E)\, E\, dE\,   r dr,
\label{eq-E_a}
\end{eqnarray}
and the total  axion luminosity is  
$L_a=4\pi D_{\rm AU}^2 \Phi_a$.

\begin{figure}
	\centering 
	\includegraphics[scale=0.40]{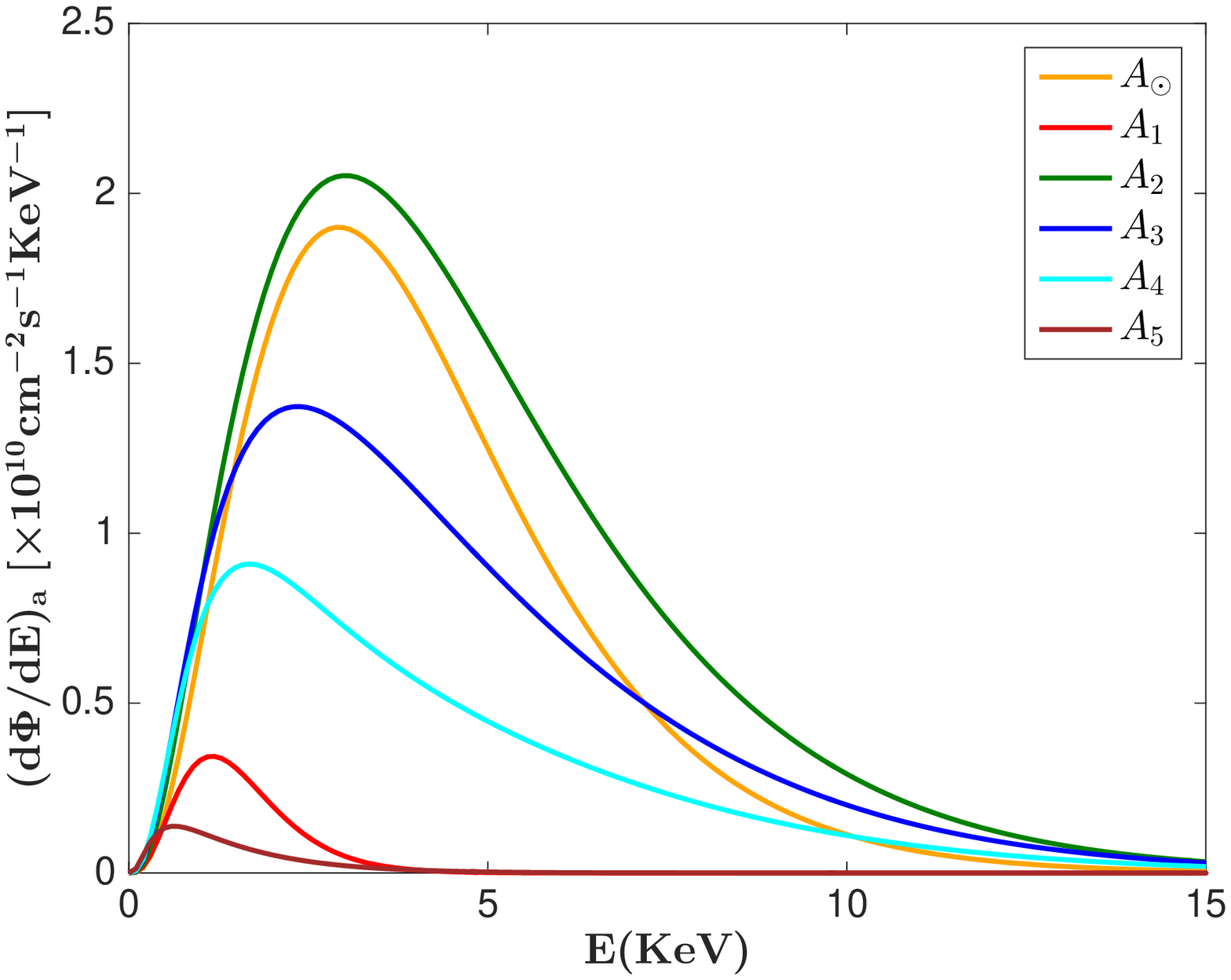}	 
	\includegraphics[scale=0.40]{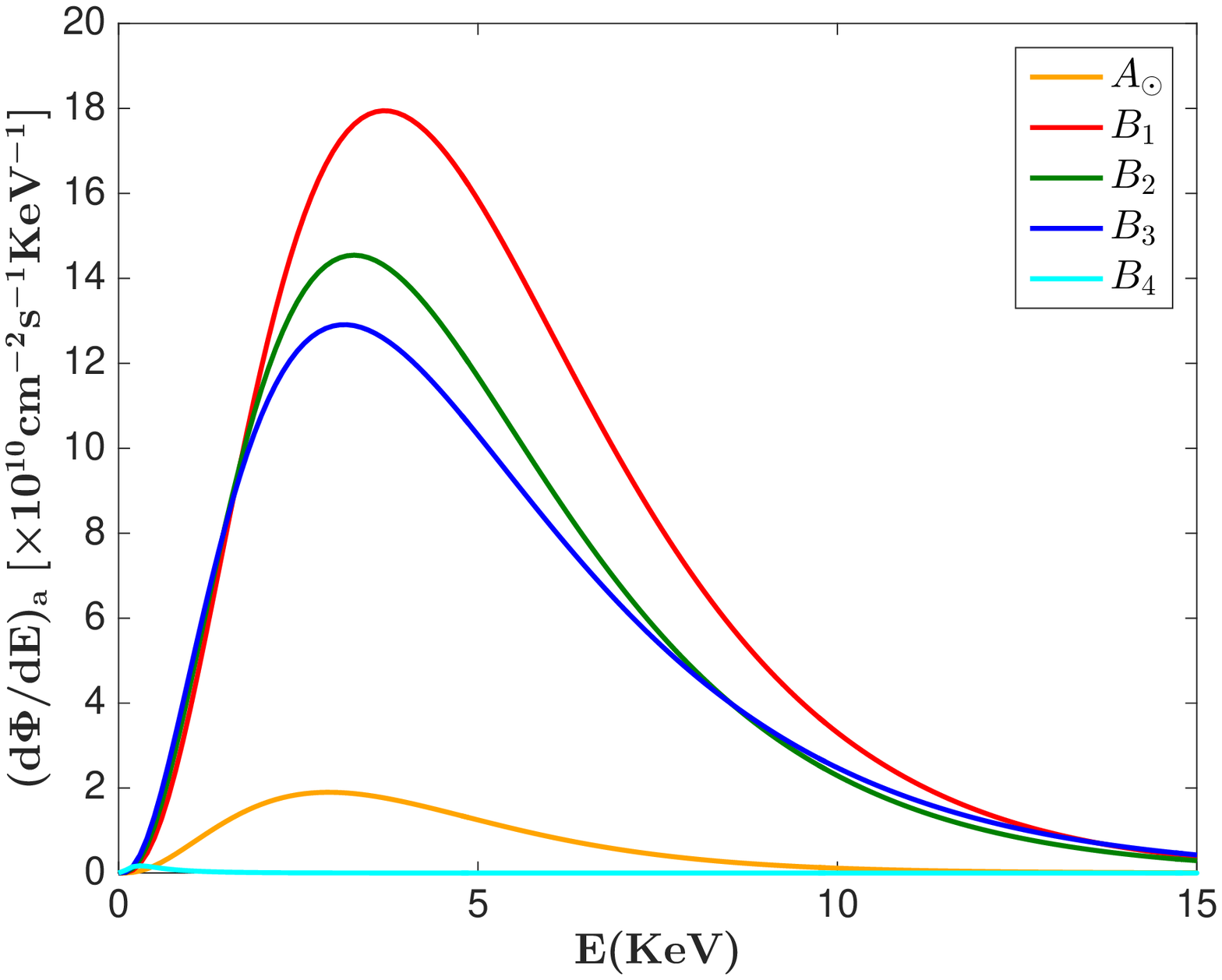}	
	\caption{Axion emission spectrum $(d\Phi/dE)_a$ due to Primakoff axion production:  We compare the axion emission fluxes for stars with 1 and 2 solar masses obtained from up-to-date stellar evolution models. The different curves correspond
		to models shown in Table \ref{tab:axionstars}:
		{\bf top panel}:
		$A_1$ ($0.01$ Gyr, red curve);  
		$A_2$  ($10.61$ Gyr, green curve);  $A_3$ ($11.37$ Gyr, blue curve);  
		$A_4$ ($11.58$ Gyr, cyan curve);  $A_5$ ($12.16$ Gyr, brown curve); 
		{\bf lower panel}:
		$B_1$ ($0.02$ Gyr, red curve);$B_2$ ($0.7$ Gyr, gree curve);
		$B_3$ ($0.98$ Gyr, blue curve); $B_4$ ($1.05$ Gyr, cyan curve);	
		The {\bf Sun} ($A_\odot$,  $4.6$ Gyr, orange curve) is plotted in both panels. For the calculation of the axion spectrum, we use an axion–photon coupling of $g_{a\gamma}=5\times 10^{-11}\; {\rm GeV^{-1}}$.} 
	\vspace{0.2cm}
	\label{fig:Primakoff_inside}
\end{figure}

\medskip\noindent
In Table \ref{tab:axionstars} we summarized  the main characteristic of the stellar models considered in this study. Figure \ref{fig:Primakoff_inside_r} shows the contour plot of the axion luminosity inside the star [Eq. (\ref{eq:dPhidE})] as a function of  the axion energy E and the dimensionless radius for some stellar models. Figure \ref{fig:Primakoff_inside} shows the axion emission spectra of the same stellar models present in Table \ref{tab:axionstars}.  For convenience we consider that all stars are located at a fiducial  distance of 1 A.U.

\medskip\noindent
The stellar models were computed with  the release version 12115 of the stellar evolution code MESA \citep{2011ApJS..192....3P,2013ApJS..208....4P,2015ApJS..220...15P,2018ApJS..234...34P,2019arXiv190301426P}. For the computation of these stellar models, we assume the input physics used for the standard solar model (SSM) but with the necessary adaptation for these stars. The SSM is a reference solar model: a one-dimensional
 stellar evolution of one solar mass star  allowed to evolve in time until the present-day solar age, $4.57$ Gyrs, having been calibrated to the values of luminosity and effective temperature of the present Sun, respectively, $ 3.8418 \times 10^{33}$ erg s$^{-1}$ and $ 5777$ K, as well as the observed abundance ratio at the Sun's surface: (Z$_\text{s}$/X$_\text{s}$)$_{\odot}=0.0181$, where $Z_s$ and $X_s$ are the metal and hydrogen abundances  at the surface of the star \citep{1993ApJ...408..347T,1995RvMP...67..781B,2006ApJS..165..400B}.  
 The details of this SSM in which we use the AGSS09 (low-Z) solar abundances \citep{2009ARA&A..47..481A}  can be found in \citet{2013MNRAS.435.2109L} and \citet{ 2020MNRAS.498.1992C}.  All the generated one and two-solar mass models start in  the pre-main sequence, assuming the star was initially chemically homogeneous and fully convective. 

\medskip\noindent
The SSM corresponds to the model $A_\odot$ in Table \ref{tab:axionstars}. In Figs. \ref{fig:Primakoff_inside_r} (top-left panel) and  \ref{fig:Primakoff_inside} (top and lower panels -- blue light curve) is shown the  solar axion luminosity  contours and the axion solar spectrum, respectively. In the first line of columns 7 and 8 of Table \ref{tab:axionstars} are shown some characteristics of the axion emissivity for the Sun. In particular, we check that for an axion model with $g_{a\gamma}=10^{-10}\;{GeV}^{-1}$, the total solar axion flux at the Earth is $3.75\;10^{11}\;{\rm cm^{-2} s^{-1}}$ near the value predicted originally  by  \citet{1989PhRvD..39.2089V},  which corresponds to  a luminosity  $1.9\;10^{-3} L_\odot$. The properties of this axion model are consistent with the axion emissivity contours and  emission spectrum found in the literature for the Sun, for instance by \citet{2007JCAP...04..010A} and 
\citet{2009JCAP...02..008A}, respectively.

\medskip\noindent
Table \ref{tab:axionstars} shows five solar-mass star models at different stages of evolution from the pre-main sequence up to the red giant phase, and another four models for two solar-mass stars  at identical evolution stages.
Conveniently, we choose the fiducial value of $g_{a\gamma}=5\;10^{-11}\;{\rm GeV}^{-1}$; this value is slightly lower than the current upper limit found by CAST helioscope \citep{2017NatPh..13..584A}. 
Figure \ref{fig:Primakoff_inside_r} also shows that the axion emission always occurs in the stellar core, within a radius smaller than $0.5 R_\star$.    
The magnitude and energy range of the axions emitted is strongly dependent on the physics of the stellar core [cf. Eqs. (\ref{eq:Gamma_axion}) and (\ref{eq:k_axion})]. Specifiably, the core's  temperature and density for these stars varies from $5.2\,10^6$ to $2.5\,10^7$ $^oK$ and from $3$ to $1.5\,10^3$  $g\;cm^{-3}$. 

This dependence of the axion emissivity on the structure of the star leads to quite distinct axion spectra shapes in comparison with the Sun (see  Fig. \ref{fig:Primakoff_inside}).
For instance, one of the two solar-mass stars has an axion spectrum  that is 15 times larger than the solar one (compare $A_\odot$ and $B_1$ models in Fig. \ref{fig:Primakoff_inside}).   Moreover, the average  axion energy $\left\langle E_{a}\right\rangle$   and axion luminosity  $L_a$ vary significantly between models  (see Table \ref{tab:axionstars}). Equally, the axion luminosity changes by several orders of magnitude from  $10^{-3}$ to $10^{-6} L_\odot$ (see Table \ref{tab:axionstars}),
and the shape of the axion spectra changes slightly with the star; the more massive stars have a more intense axion emission spectrum as shown in Fig. \ref{fig:Primakoff_inside}.

\begin{table}
	\raggedleft
	\begin{tabular}{lllllllll}
		{\small Star} & {\small  Mass} & {\small Age} &  {\small Radius} & {\small Lum.} & {\small Temp.}  
		& $\left\langle E_a\right\rangle$ & $L_a$  & $L_{a\gamma}$  \\
		&  $M_\odot$ & Gyr &  $R_\odot$ & $L_\odot$  & {\small $10^3 {\rm ^oK}$}  &    ${\rm KeV}$  &  $L_\odot$ & $L_\odot $ \\
		\hline
		\\
		$A_{\odot}$   &  $1.0$  &   $4.6$   & $1.0$    & $1.0$  & $ 5.87 $  &  $4.2$ &  $4.6\; 10^{-4}$   &  $3.7\; 10^{-7}$ \\
		$A^{*}_{\odot}$   &  &     &    &   &   & $4.2$  & $1.8\; 10^{-5}$    &  $5.9\; 10^{-10}$ \\
		$A^{**}_{\odot}$   &  &     &    &   &   &   &     &  $4.9\; 10^{-8}$ 
		\\ \\
		$A_{1}$   &  $1.0$  &   $0.01$   & $1.41$    & $0.59$  & $ 4.27 $  &  $1.6$ & 		$1.1\; 10^{-5}$   &  $1.6\; 10^{-8}$  \\		
		$A_{2}$   &  $1.0$  &   $10.6$   & $1.45$    & $1.92$  & $ 5.64$  &  $4.8$ &  $6.8\; 10^{-4}$   &  $1.1\; 10^{-6}$  \\
		$A_{3}$   &  $1.0$  &   $11.4$   & $1.75$    & $2.10$  & $ 5.3 $  &   $4.7$ &  $4.4\; 10^{-4}$   &  $1.1\; 10^{-6}$  \\
		$A_{4}$   &  $1.0$  &   $11.6$   & $2.01$    & $2.25$  & $ 5.0$   & $4.4$ & 	$2.4\; 10^{-6}$   &  $1.1\; 10^{-7}$  \\
		$A_{5}$   &  $1.0$  &   $12.2$   & $6.0$    & $15.3$  & $ 4.7 $  &    $1.5$ &  $4.2\; 10^{-6}$   &  $2.1\; 10^{-7}$  \\
		\\ \\
		$B_{1}$   &  $2.0$  &   $0.02$   & $1.63$    & $16.1$  & $ 9.1$  &   $5.3$ &   $6.9\; 10^{-3}$   &  $1.4\; 10^{-5}$  \\
		$B^{*}_{1}$   &  &     &    &   &   & $5.3$  & $2.8\; 10^{-4}$    &  $2.2\; 10^{-8}$ \\
		$B^{**}_{1}$   &  &     &    &   &   &   &     &  $1.0\; 10^{-6}$ 
		\\\\
		$B_{2}$   &  $2.0$  &   $0.7$   & $2.42$    & $19.9$  & $ 7.84$  &   $5.0$ &   $5.1\; 10^{-3}$   &  $2.3\; 10^{-5}$  \\ 
		$B_{3}$   &  $2.0$  &   $1.0$   & $3.50$    & $28.8$  & $ 7.15$  &  $5.2$ &   $1.1\; 10^{-6}$   &  $2.1\; 10^{-7}$  \\ 
		$B_{4}$   &  $2.0$  &   $1.1$   & $16.2$   & $101.9$  & $ 4.56$  &  $0.7$ & $1.1\;10^{-6}$   &  $2.0\;10^{-7}$    \\ 
		\\
		\hline
	\end{tabular}
	\caption{Comparison of the axion properties of several stellar models, including  an up-to-date standard solar model of the Sun \citep{ 2020MNRAS.498.1992C}. The table shows different stellar quantities such Mass, Age, Radius, Luminosity (Lum.), Temperature (Tem.), Axion Luminosity  and
		x-ray axion Luminosity (see the main text).	
		The axion has a mass of $m_a=10^{-7}\;{\rm eV}$ and  axion–photon coupling $g_{a\gamma}=5\;10^{-11}\;{\rm GeV^{-1}}$. The magnetic conversion layer in the star has a thickness of  $\Delta R_\star=0.15R_\star$  and an average magnetic field of $B=3\;10^4\;{G}$. In the calculation of the model pairs: $A^{*}_{\odot}$ and  $B^{*}_{1}$, and $A^{**}_{\odot}$ and  $B^{**}_{1}$,  we assume  $g_{a\gamma}=10^{-11}{\rm GeV}^{-1}$, and  $m_a=10^{-5}{\rm eV}$ respectively.
		\label{tab:axionstars}}
	\vspace{0.5cm}
\end{table}

\section{V. Inverse Primakoff  interaction}

In some regions of the atmosphere of most stars, the stellar plasma has the conditions to  naturally stimulate the conversion of axions to photons \citep{2009NJPh...11j5020Z}.  The inverse Primakoff reaction occurs when the energy of the photons is sufficiently far above all resonances [Eq. \ref{eq:Kappa}]. Since we are concerned with 
axions with energy in the KeV range this condition automatically satisfies \citep{1989PhRvD..39.2089V}. Typically, such an axion process occurs in a region made of a plasma of low density and metallicity immersed in a strong magnetic field.
To that end, we will assume that the inverse Primakoff reaction occurs in a magnetized region of the star with a low density. 
This condition is satisfied in the outer layers of most main-sequence and post-main sequence stars since most of the stellar mass locates in the internal region of the star. For instance for a star like the Sun, 98\% of its mass is below $0.7$ of its radius \citep{ 2020MNRAS.498.1992C}. The external layer of these stars with very low densities are known to be responsible for the generation of a strong magnetic field by dynamo action when in the presence of differential rotation and meridional flows \citep{2014SSRv..186..535L}. Accordingly, as predicted by many stellar dynamo models, we assume the existence of a magnetic layer located between the upper layers of the star and lower part of the stellar atmosphere, where the near-surface convection influences the local magnetic fields. This most external layer is responsible for the magnetic activity observed in many stars \citep[e.g.,][]{2008ApJ...686.1420P,2015A&A...581A..42B,2017RSOS....460271B}.  Such large concentrations of a magnetic field emerging on the stellar surface lead to the formation of active regions encompassing magnetic features, such as dark spots and bright faculae \citep{2006RPPh...69..563S,2020A&A...633A..32S}.

\medskip\noindent
Under the approximation that
all axions travel along radial trajectories in the stellar
atmosphere, and the hypothesis that the stellar magnetic field
varies in a length scale much larger than the photon and axion wavelengths, we can derive an analytic expression 
for the axion conversion to photons using a WKB approximation  \citep{1988PhRvD..37.1237R}. Accordingly,  the propagation of the photon and axion in the radial direction with a Energy $E$ is given by
\begin{eqnarray} 
\left[i\partial_r+E+ 
\mathbf{\hat{A}}\right]
\cdot \mathbf{\hat{V}}=\mathbf{0},
\label{eq-3dwave}
\end{eqnarray} 
where  $\mathbf{\hat{V}}$ is a 3 component vector and $\mathbf{\hat{A}}$ 
is a $3\times 3$ symmetric matrix  that defines the interaction of the photon-axion pair with the magnetic field $\mathbf{B}$.  The vector $\mathbf{\hat{V}}$ reads
\begin{eqnarray} 
\mathbf{\hat{V}}=
\begin{pmatrix}
V_{\perp}\\
V_\parallel \\
V_a \\
\end{pmatrix},
\label{eq-3dwaveV}
\end{eqnarray} 
where $V_\parallel (r)$ and $V_{\perp}(r)$  denotes the vector potential in the plane normal and perpendicular  to the direction of the propagation and parallel to the external magnetic field, while $V_a(r)$ is the axion field.
The matrix  $\mathbf{A}$ reads
\begin{eqnarray} 
\mathbf{\hat{A}}=
\begin{pmatrix}
\Delta_{\perp} & 0 & 0\\
0& \Delta_\parallel & \Delta_{B}   \\
0& \Delta_{B}   & \Delta_{a}  \\
\end{pmatrix}.
\label{eq-3dwaveA}
\end{eqnarray}  
Here, we follow the notation of \citet{1988PhRvD..37.1237R}.
The quantities $\Delta_{\perp}(r)=4/2E\;\xi \sin^2{\phi}$,
$\Delta_\parallel(r)=7/2\;E\xi \sin^2{\phi}$ and
 $\Delta_B(r)=(g_{a\gamma}/2) B(r) \sin{\phi}$ 
are terms that define the interaction of the axion with 
the magnetic field, and $\Delta_a=-m_a^2/(2E)$ is a term that incorporates the axion mass $m_a$ and is responsible for the small difference momenta between axion and photon states. $B(r)$  is the strength of the magnetic field at radius $r$ and $\phi$ is the angle between  the magnetic field and the photon momentum, and $\xi(r)=(\alpha_{em}/45\pi)[B(r)/B_{c}]^2$ where  
$B_c=m_e^2/e$ is a critical magnetic  field strength.  

\medskip\noindent
In the following, we discuss the conversion of photon to axion mixing as a photon propagates through a highly magnetized medium.  
Following \citet{1988PhRvD..37.1237R}, if the magnetic field is homogeneous, the subsystem of equations  $(V_\parallel, V_a)$ on the system of equations (\ref{eq-3dwave}) [also the second and third lines of (\ref{eq-3dwaveV}) and (\ref{eq-3dwaveA})] can be computed separately from the  equation $V_{\perp}$ (also the first line of  \ref{eq-3dwaveV} and \ref{eq-3dwaveA}). Therefore,  the leading stationary wave Eq. (\ref{eq-3dwave})  reduces to two decoupled systems of first-order differential equations \citep[e.g.,][]{2019JCAP...11..020F,2012ApJ...748..116P}.
Accordingly,  the subsystem of Eqs.  $(V_\parallel, V_a)$ reduces to the simplified stationary  wave equation
\begin{eqnarray} 
\left[i\partial_r+E+ 
\mathbf{A}\right]
\cdot \mathbf{V}=\mathbf{0}
\label{eq-2dwave}
\end{eqnarray} 
where  $\mathbf{V}$  is a vector (similar $\mathbf{\hat{V}}$)
with only two component that defines the  amplitudes of photon and axion states, $\mathbf{A}$ is a $2\times 2$ symmetric matrix 
(similar $\mathbf{\hat{A}}$) that defines the interaction of the axion
with the magnetic field. Accordingly,  $\mathbf{V}$ reads
\begin{eqnarray} 
\mathbf{V}=
\begin{pmatrix}
V_\parallel \\
V_a \\
\end{pmatrix}
\end{eqnarray} 
and  $\mathbf{A}$ takes the form
\begin{eqnarray} 
\mathbf{A}=
\begin{pmatrix}
\Delta_\parallel & \Delta_{B}   \\
\Delta_{B}   & \Delta_{a}  \\
\end{pmatrix}.
\end{eqnarray} 

\medskip\noindent
\citet{2019JCAP...11..020F} have made a detailed study of the conversion  of axionlike particles to photons in a electromagnetic background have shown that Eq. (\ref{eq-2dwave}) is valid also in the high-magnetization limit where the electron cyclotron frequency $\omega_c=\sqrt{\alpha_e}B/(m_e c)$ is much larger than electron plasma frequency $\omega_{pl}$ [see Eq. \ref{eq:omegapl}]  and  $E$ ($\omega_c\gg E$ and $\omega_{pl} $).
Equation (\ref{eq-2dwave}) can be diagonalized by  rotating  the fields, $\mathbf{V^\prime}=\mathbf{R(-\theta)}\cdot\mathbf{V}$, 
where $\mathbf{R(\theta)}$  is the two dimensional unitary rotation matrix.  $\mathbf{R(\theta)}$  is given by
\begin{eqnarray} 
\begin{pmatrix}
\cos{\theta} & -\sin{\theta}  \\
 \sin{\theta}   & \cos{\theta}  \\
\end{pmatrix},
\end{eqnarray} 
where $\theta$ is  a mix angle  satisfying the 
relation 
\begin{eqnarray} 
\tan{(2\theta)}=\frac{2\Delta_B}{\Delta_\parallel-\Delta_a}. 
\label{eq:theta}
\end{eqnarray} 
Accordingly, the wave  Eq. (\ref{eq-2dwave})  becomes diagonal in the $\prime$ 
referential of the rotation vector $\mathbf{V^\prime}$,
thus
\begin{eqnarray} 
\left[i\partial_r+E+ 
\mathbf{A^\prime}\right]
\cdot \mathbf{V^\prime}=\mathbf{0}.
\label{eq-2dwave.b}
\end{eqnarray} 
Here the matrix $\mathbf{A^\prime}$ reads
\begin{eqnarray} 
\mathbf{A^\prime}=
\begin{pmatrix}
\Delta_\parallel^\prime& 0   \\
0   & \Delta_a^\prime  \\
\end{pmatrix}.
\label{eq-Aprime}
\end{eqnarray} 
where $\Delta_\parallel^\prime=\lambda_1$ 
and $\Delta_a^\prime=\lambda_2$ with
$\lambda_j=1/2\left[\Delta_\parallel+\Delta_a-(-1)^j(\Delta_\parallel-\Delta_a)/\cos{(2\theta)}\right]$
($j=1,2$). 

\medskip\noindent
Following \citet{1988PhRvD..37.1237R} the previous calculation is greatly simplified if we define phases relative to the unmixed 
component, and neglect a common phase.
The solution can be found by first performing a matrix rotation to an eigenstate basis where the propagation matrix is diagonal, propagating the two eigenstates independently, and then rotating back to the photon-axion basis. This yields an evolution equation for the mixing components  in the original referential as  
\begin{eqnarray} 
\begin{pmatrix}
V_\parallel\\
V_a \\
\end{pmatrix}(r)
= 
{\cal M}(r)
\cdot\begin{pmatrix}
V_\parallel \\
V_a \\
\end{pmatrix}(0)
\end{eqnarray} 
where ${\cal M} $ is
\begin{eqnarray} 
{\cal M} 
= 
\mathbf{R(\theta)}
\cdot
{\cal M}_{D}\cdot
\mathbf{R(-\theta)}
\label{eq:M}
\end{eqnarray} 
and ${\cal M}_{D} $ is given by 
\begin{eqnarray} 
{\cal M}_{D}
= \begin{pmatrix}
e^{-i(\Delta_\parallel^\prime-\Delta_\parallel) r} & 0  \\
0   & e^{-i(\Delta_a^\prime-\Delta_\parallel) r}   \\
\end{pmatrix}.
\end{eqnarray} 

The probability for an axion-photon  
transition amplitude \citep{1989PhRvD..39.2089V,2018PrPNP.102...89I} is computed from
the off-diagonal terms in ${\cal M}_{(12)}$
 of Eq. (\ref{eq:M}). 
Therefore, the probability  of an axion being converted to a photon is 
\begin{eqnarray}
P_{a\gamma}
=  \; \sin^2{(2\theta)} \;
\sin^2{\left[\frac{\Delta_Br}{\sin{(2\theta)}}\right]}. 
\label{eq:Pagamma}
\end{eqnarray} 

The strength and geometry of the magnetic field in the atmosphere of a main sequence and post main sequence star varies in a complicated manner, in such a way that the extraction of the averaged poloidal and toroidal components of $B$ is quite diverse between these stars. Moreover, it was found that the degree of complexity of the magnetic field appears to be independent of the stellar mass. For instance in these stars the geometry surface magnetic fields varies from purely poloidal  to non-poloidal (complicated) geometries \citep{2019A&A...621A..47K}. In principle, we can consider more complex magnetic field configurations to compute $P_{a\gamma}$, 
but the estimation obtained is largely identical to the one computed with Eq. (\ref{eq:Pagamma}).  Moreover, to simplify the calculation of the x-ray spectrum produced by this inverse Primakoff reaction, we will assume the process occurs in the presence of an averaged magnetic field $B$ located in a fiducial layer of thickness (of a fixed percentage  of the stellar radius)  $\Delta R_\star$ in the exterior of star.

\medskip\noindent
In real stars, the axion magnetic conversion layers will have a very complex structure, with varying density, chemical composition and magnetic fields. Nevertheless, our model allows us to obtain for the first time an estimation of the excess of x-ray radiation due to inverse Primakoff reaction. Therefore,  we assume the magnetic layer has a thickness of $\Delta R_\star=0.15$ and is crossed by constant magnetic field $B$. 

\medskip\noindent
As we will find out in the next section, the electromagnetic emission coming from these axions will appear as an additional photon source in the x-ray energy range of the electromagnetic spectrum above the stellar background. 

\medskip\noindent
Under the previous assumptions, and since we are interested in the weak-mixing axion regime where  $\theta \ll 1$, from Eq. (\ref{eq:theta}) we compute $ \theta\approx\Delta_B/(\Delta_\parallel-\Delta_a)$, and by using this result on Eq. (\ref{eq:Pagamma}) we obtain	
\begin{eqnarray}
 P_{a \gamma}
\approx 4\frac{\Delta_B^2}{\Delta^2_a}
\,\sin{\left(\frac{\Delta_a\Delta R_\star}{2}\right)}
\label{eq:Pagamma1_approx}
\end{eqnarray}
with 
$\Delta_B^2/\Delta^2_a=g_{a\gamma}^2 B^2 E^2/m_a^4\;\sin{\phi}$.
The derivation of the previous expression is possible, since for axions with a very  low mass, we have  $|\Delta_a|\gg \Delta_\parallel$ and  $ \theta\approx\Delta_B/\Delta_a$.

\begin{figure}
	\centering 
	\includegraphics[scale=0.45]{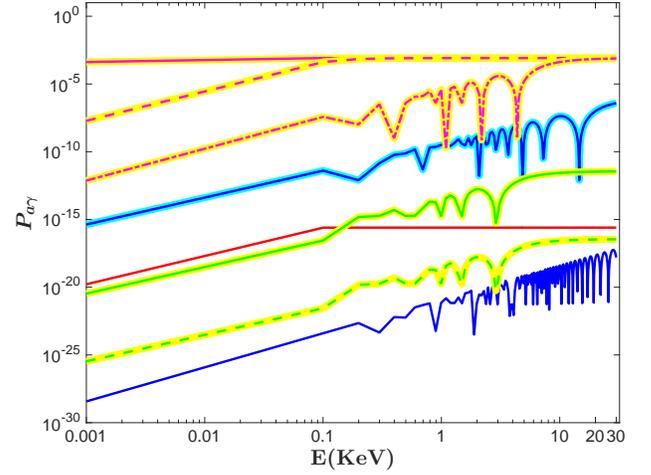}	 
	\caption{The conversion probability of an axion to photon $P_{a \gamma}$ as a function of the energy. The $P_{a \gamma}$ is obtained by varying some of the parameters of the reference axion model (see end of this caption):
		{\bf (a)} red and yellow curves correspond to Eqs. (\ref{eq:Pagamma}) and (\ref{eq:Pagamma1_approx})  with  $m_a=10^{-7}\; {\rm eV}$ (continuous curve), $10^{-6}\; {\rm eV}$ (dashed curve) and $10^{-5}\; {\rm eV}$ (dotted-dashed curve);
		{\bf (b)} blue and cyan curves  correspond to an axion model identical to {\it (a)} with $m_a=10^{-4}\; {\rm eV}$ and  $\Delta R_\star=0.005\; R_\odot $;
		{\bf (c)} green and yellow curves correspond to an axion model identical to {\it (a)} with $m_a=10^{-3}\; {\rm eV}$ and $\Delta R_\star=0.001\; R_\odot $;
		{\bf (d)} dashed green and yellow curves  correspond to a axion model identical to {\it (c)} with $\phi=10^{-3}\pi$;
		{\bf Reference axion model} $g_{a\gamma}=5\times 10^{-11}\;{\rm GeV^{-1}}$,  $m_a=10^{-7}\;{\rm GeV}$,  $B=3\times10^{4}\;{G}$, $\phi=\pi/2$ and $\Delta R_\star=0.15\; R_\star $ (see main text). 
		For comparison, we  also show $P_{a\gamma}$ for the conversion of axions to photons occurring inside an experimental axion detector (with a magnetic field $B$ and dimension $D_m$): (e) red curve corresponds to an axion with  $m_a=1\times 10^{-3}\; {\rm eV}$, and a detector with  $B=9\times 10^{4}\;{G}$ and $D_{\rm m} =10\;{\rm m}$ \citep{2009NJPh...11j5020Z};	 (f)  blue curve corresponds to conversion to an axion with $m_a=8\times 10^{-4}\; {\rm eV}$ and a detector with  $B= 3\times 10^{-1}\;{G}$ and  $D_{\rm m} =6\times 10^{5}\;{\rm m}$; $D_{\rm m}$  is identical to $ \Delta R_\star$  for a star (see main text)\citep{2014MNRAS.445.2146F,1983PhRvL..51.1415S}.
	} 
	\label{fig:Pagamma}
\end{figure}

\medskip\noindent
Figure \ref{fig:Pagamma} shows Eqs. (\ref{eq:Pagamma}) and (\ref{eq:Pagamma1_approx})  in comparison with other expressions in the literature. We start by noticing that for all  parameters considered in this study, the two expressions are equivalent. 
In this figure, we choose a set of axion parameters that better shows the similitude of the inverse Primakoff reaction occurring in the external layers of stars and on axion detectors. 
The numerical results obtained with our analytical expressions are identical to others found in the literature to determine the conversion of axions to photons in experimental detectors, for instance \citet{2009NJPh...11j5020Z} and \citet{1983PhRvL..51.1415S}.
$P_{a \gamma}$ [see Eq. \ref{eq:Pagamma1_approx}] function is strongly dependent on the axion parameters (mass, energy and axion-photon coupling constant) and the magnitude of the magnetic field, as shown by the approximative Eq. (\ref{eq:Pagamma1_approx}) and confirmed in Fig. \ref{fig:Pagamma}.
Some of $P_{a\gamma}$ results show a oscillatory pattern 
(see cases {\bf (a) },  {\bf (b) } and  {\bf (c) } in Fig. \ref{fig:Pagamma}), as well as the $P_{a\gamma}$  of \citet{2009NJPh...11j5020Z}. This behavior is related to
the mass of the axion $m_a$ and  $\Delta R_\star$ (see Eq. \ref{eq:Pagamma1_approx}). 
In an axion detector, $D_{\rm m}$  replaces $\Delta R_\star$, where $D_{\rm m}$  is the dimension of the region where $B$ is applied \citet{2009NJPh...11j5020Z}. 
We also observe that  $P_{a\gamma}$  depends on the phase $\phi$ as shown by the analytical approximation (\ref{eq:Pagamma1_approx}), leading in some cases to the $P_{a\gamma}$ functions varying by a few orders of magnitude (see cases {\bf (c)} and {\bf (d)}  in Fig. \ref{fig:Pagamma}).  Nevertheless, without loss of generality, we opt to choose $\phi=\pi/2$ or $\sin{\phi}=1$ if not stated otherwise.

 \section{VI. Axion electromagnetic inverse Primakoff spectrum}

Here, we calculate the electromagnetic spectrum related with the inverse Primakoff reaction. Such spectrum is obtained as the product of the axion emission spectrum coming from the stellar interior [Eq. \ref{eq:dPhidE}] and the axion-to-photon conversion probability $ P_{a\gamma}$ [Eq. \ref{eq:Pagamma}]. Accordingly,  the flux of axion-induced x-ray photons at 1 A.U. is calculated as
 \begin{eqnarray}
 \left(\frac{d\Phi}{dE} \right)_{a\gamma}=
P_{a\gamma}\;
 \left(\frac{d\Phi}{dE} \right)_{a},
 \label{eq:dPhidE_xray}
 \end{eqnarray}
where $({d\Phi}/{dE})_{a}$ is given by Eq. (\ref{eq:dPhidE}) for $r=R_\star$.
 
\medskip\noindent
 The total axion flux conversion to  x-ray flux at an 1 A.U. distance, the correspondent luminosity, and the mean x-ray axion energy  is computed in a similar way to Eqs. (\ref{eq-Phi_a})  and (\ref{eq-E_a}), but now modified by the  
 conversion probability $ P_{a\gamma}$ [Eq. \ref{eq:Pagamma}]. 
 Accordingly  
 \begin{eqnarray}
 \Phi_{a\gamma}=2\pi \int_0^1 \int^\infty_{\omega_{\rm pl}} \varphi_a (r,E)\, P_{a\gamma} \, dE\,   r dr,
 \end{eqnarray}
the  corresponding x-ray luminosity is $L_{a\gamma}=4\pi D_{\rm AU}^2 \Phi_{a\gamma}$. The predictions of the x-ray axion luminosities for the stars studied in this work  are shown in column 9 of Table \ref{tab:axionstars}.

\begin{figure*}
	\centering 
	\begin{tabular}{cc}
		\includegraphics[scale=0.40]{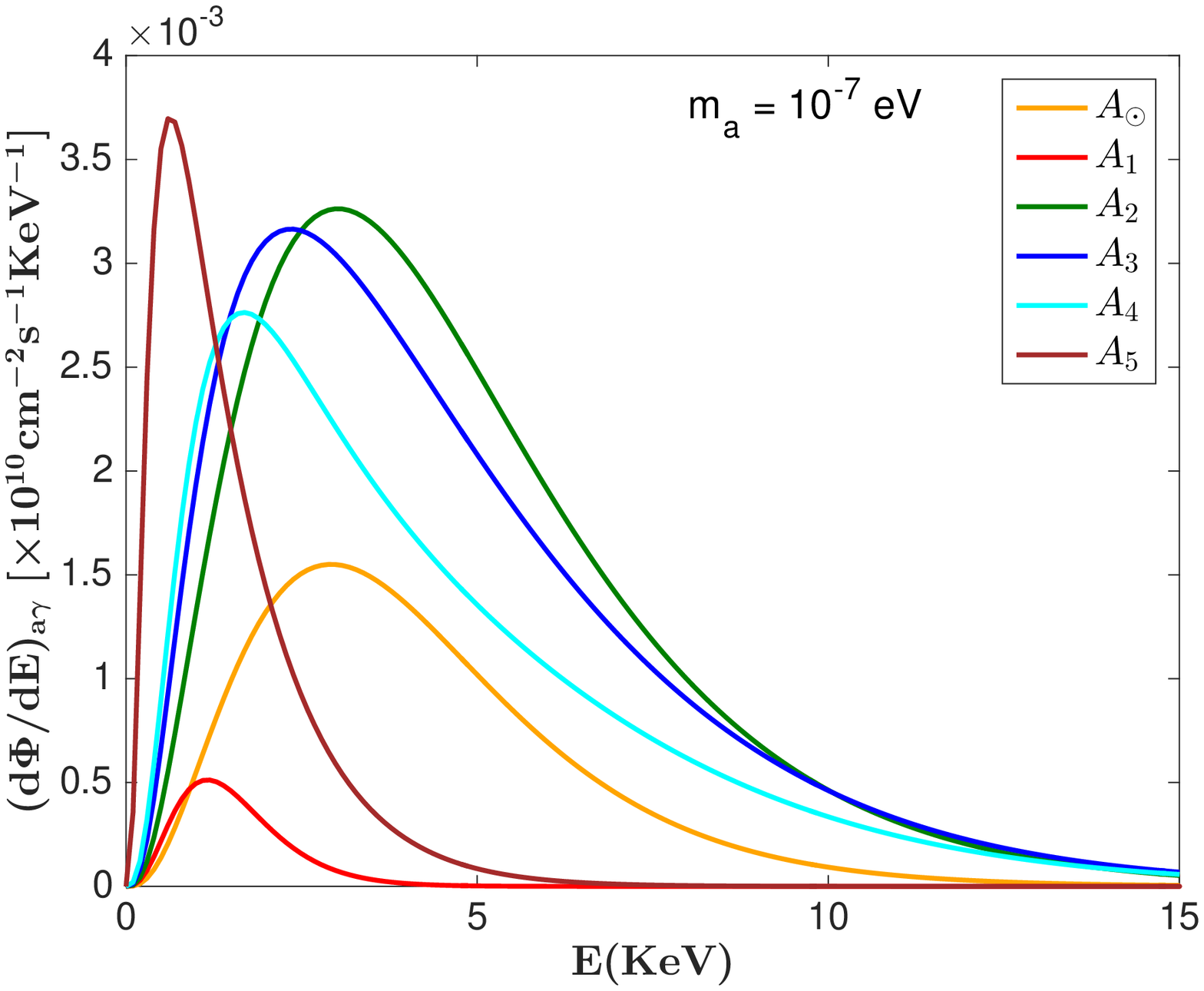}	 
		&\hspace{-0.1cm}	
		\includegraphics[scale=0.40]{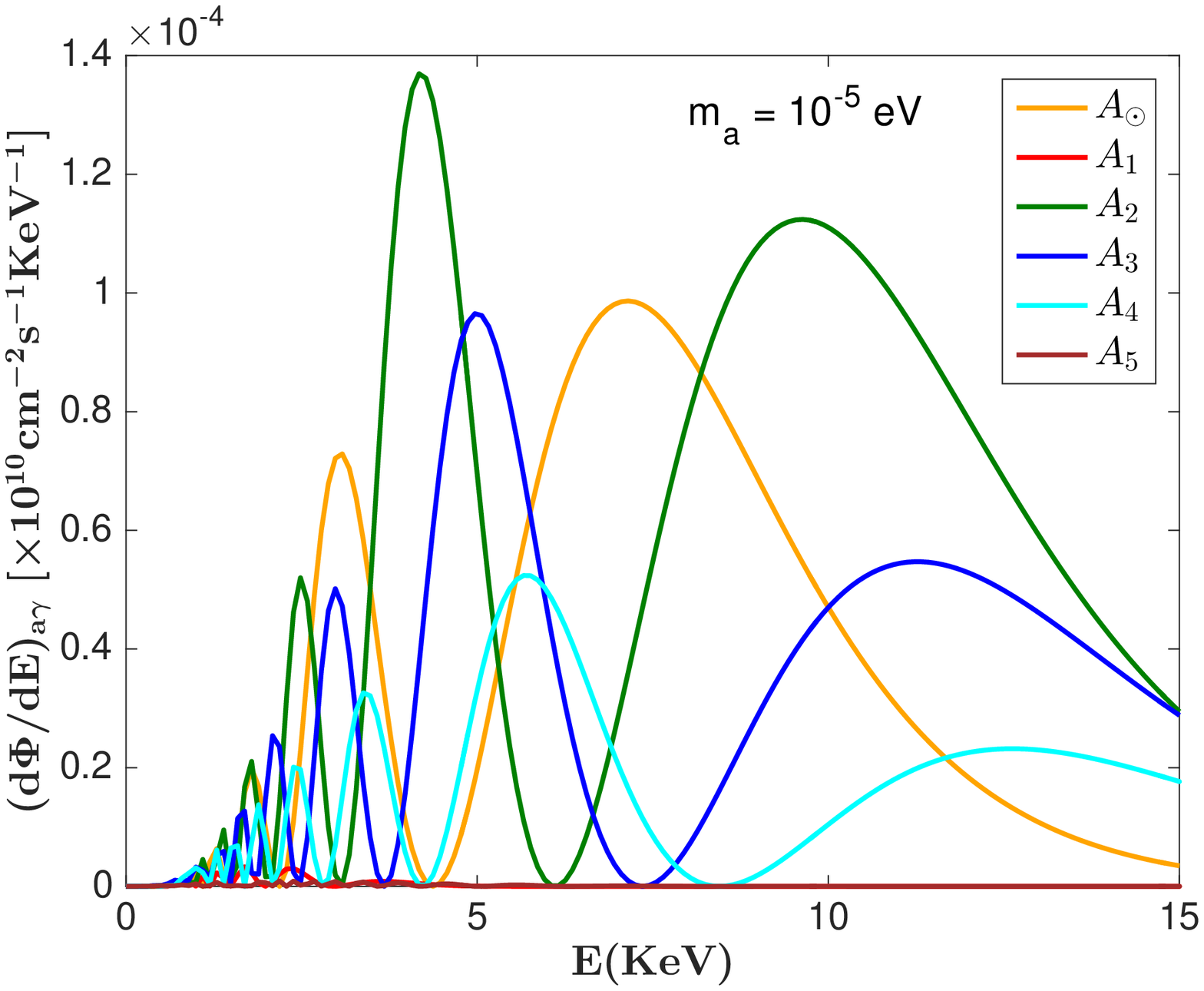}
		\\
		\vspace{-0.4cm}	
		\\
		\includegraphics[scale=0.40]{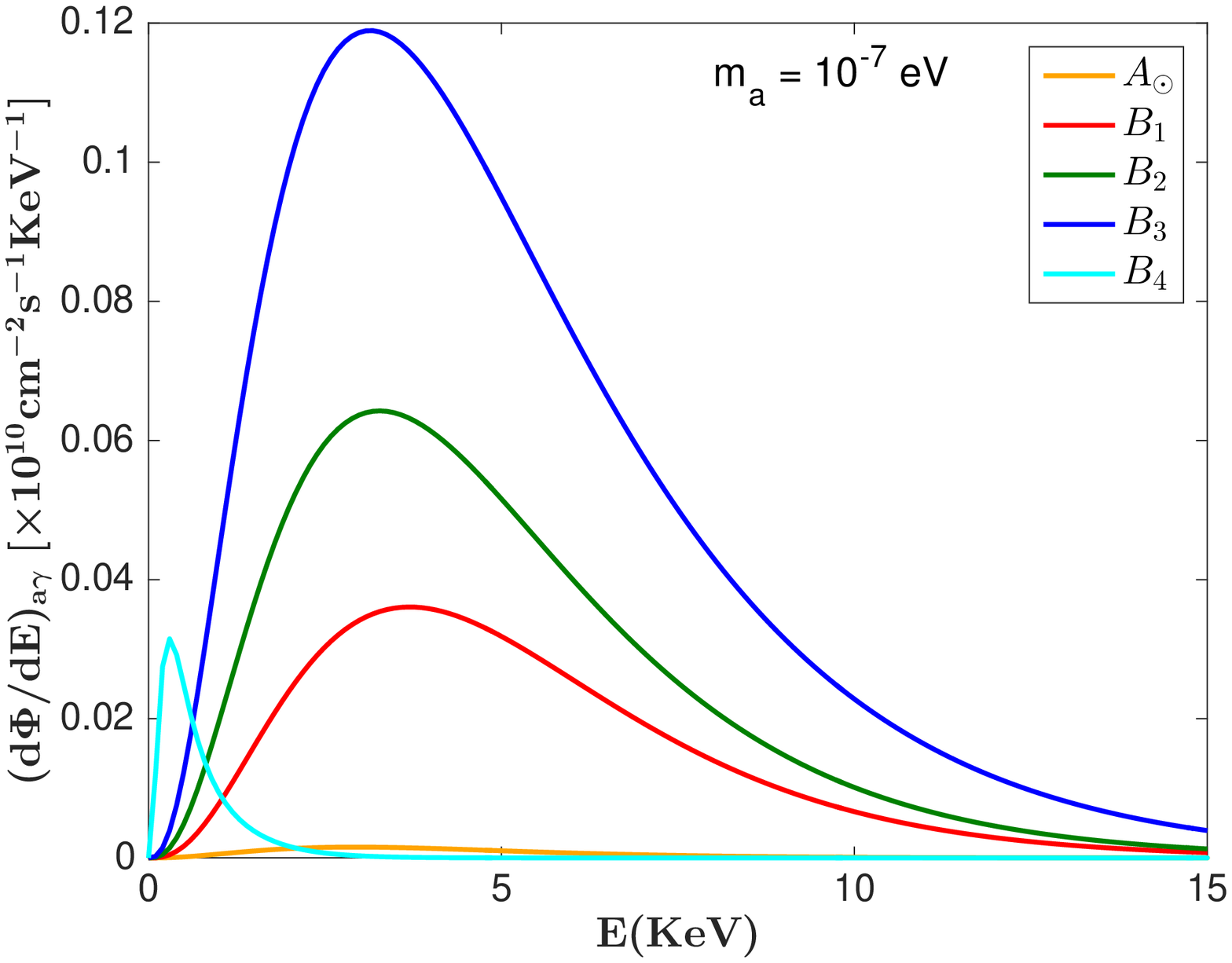}
		&\hspace{-0.1cm}	
		\includegraphics[scale=0.40]{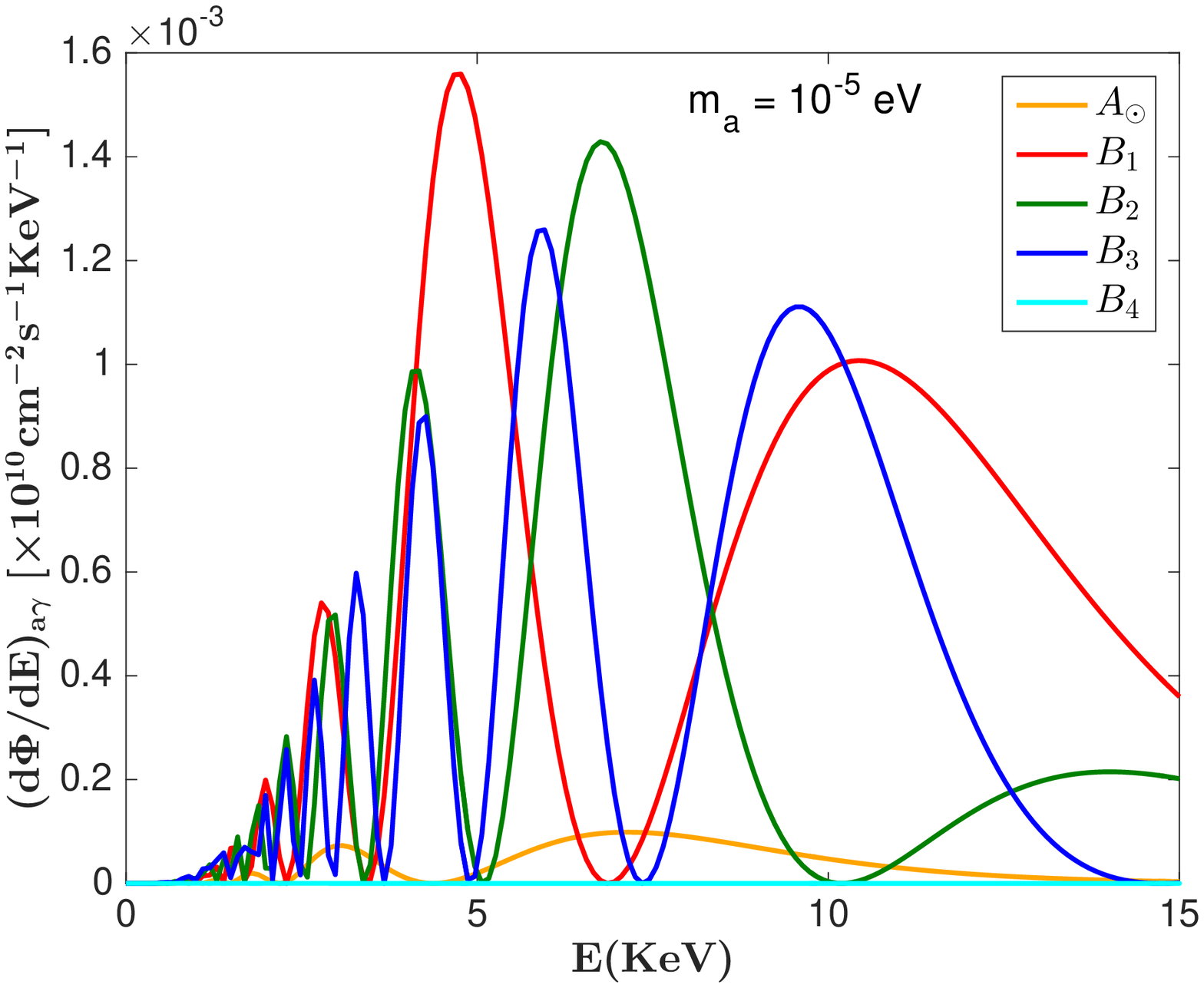} \\
	\end{tabular}	
	\caption{x-ray emission spectrum $(d\Phi/dE)_{a\gamma}$ due to inverse Primakoff photon production: Comparison spectra calculated  for stellar models of one ({\bf top panels: a, b})  and two ({\bf lower panels:  c, d})  solar masses shown in Fig. \ref{fig:Primakoff_inside} and Table \ref{tab:axionstars}.  The mass of the axion is equal to:  $10^{-7}\;{\rm eV}$ ({\bf a})  and  $10^{-5}\;{\rm eV}$  ({\bf b})  for models $A_\odot,A_1,\cdots,A_5$; and   $10^{-7}\;{\rm eV}$  ({\bf c})  and  $10^{-5}\;{\rm eV}$ ({\bf d}) for models $A_\odot,B_1,\cdots,B_4$.  In all these models the  axion magnetic conversion layer has a thickness of $0.15 \,R_\star$ and  an averaged magnetic field of $3\;10^4\;{G}$.  \label{fig:Primakoff_outside}
	} 
\end{figure*}

\medskip\noindent
Figure \ref{fig:Primakoff_outside} show the typical x-ray spectra generated by the magnetic layer located in the external layers of the stars (see Table \ref{tab:axionstars}).  All these x-ray axion spectra are a few orders of magnitude smaller that the axion spectra coming from the core of the star, accordingly the x-ray axion  luminosity varies  by the same other of magnitude (see column 9 of Table \ref{tab:axionstars}).

\medskip\noindent
We found that some of the x-ray axion spectra have a unique sinusoidal shape that is related to the mass of the axion, as illustrated in Fig. \ref{fig:Primakoff_outside} [panels  {\bf (b)}  and {\bf (d)}].  Such periodic behavior is more pronounced for axions with a lower mass, the variation of the mass of the axion leads to spectra with similar sinusoidal shapes but different periodicities. 

\medskip\noindent
Since such stars' magnetic field is time-dependent like the stellar magnetic cycle \citep{1995ApJ...438..269B,2014SSRv..186..535L}, the x-ray spectrum produced from the interaction of axions with the stellar magnetic fields (see section  \ref{sec-ASMF} and references therein) will also vary in time.

\medskip\noindent
The spectra calculated here are initial estimates of the averaged x-ray emission rate of axion conversion in the atmosphere of a star. To make reliable predictions of the x-ray emission spectra is necessary to include a more detailed account of the stellar atmosphere structure. Indeed,	 the atmospheres of low-mass stars, among others, present a large diversity in terms of thermodynamics properties, chemical composition and intensity and topology of the magnetic fields.  
Only by including a detailed description of the atmosphere of such stars, it is possible to make	a reliable prediction of the x-ray emission coming from converting axions to x-rays in stellar atmospheres.

\medskip\noindent
It is worth mentioning that axion spectra and the x-ray  spectra of the axion emission 
are strongly dependent on the axion-photon coupling constant
and the  mass of the axion. For instance, if the value of the axion-photon coupling constant  changes from  
$5\times 10^{-11}\,{\rm GeV }^{-1}$ to  $10^{-11}\,{\rm GeV }^{-1}$,
$L_a$ for the Sun and a two-solar mass star decreases by an order of magnitude,   while $L_{a\gamma}$ decreases by 3 orders of magnitude  (compare model $A^{*}_\odot$ with $A_\odot$, and $B^{*}_1$ with $B_1$ in Table \ref{tab:axionstars}). 
If similarly, when we  vary $m_a$ from $10^{-7}\;{eV}$ to $10^{-5}\;{eV}$, then $L_{a\gamma}$ decreases by one order of magnitude (compare model $A^{**}_\odot$ with $A_\odot$, and $B^{**}_1$ and $B_1$ in Table \ref{tab:axionstars}). 
 
\medskip\noindent
Nevertheless, we can find a combination of parameters, such as  $g_{a\gamma}=5 \times 10^{-11}\,{\rm GeV }^{-1}$ and $m_a=10^{-7}\,{eV}$ 
(as show in Table \ref{tab:axionstars}) for which the  x-ray axion luminosities of these stars vary between
$10^{-5}$ and  $10^{-7}\,L_\odot$.  Such luminosities  
are  comparable or even larger than the x-ray luminosity currently measured for many low-mass stars: $10^{-4}$ to $10^{-7}\; L_\odot$
\citep[][]{2016MNRAS.462.4442S}.  This result shows the potential of
x-ray stellar astronomy to put constraints in low mass axion models.

\section{VII. Conclusion} 
\label{sec-DC}

Stars are commonly used to test new particles and interaction channels, mostly because many of such interactions occurring inside stars are strongly dependent on temperature, density and chemical composition of the stellar plasma. By taking advantage of similar behaviors for the axion, we study how such particles are generated inside main-sequence and post main-sequence stars. We have confirmed previous computations done for the present Sun and found that for other low-mass stars the axion spectrum changes in intensity and shape by a few orders of magnitude, depending significantly on the mass and age of the star. 

\medskip\noindent
We found that stars with external magnetic fields could have an excess of radiation in the x-ray energy range of the electromagnetic spectrum resulting from the conversion of axions into photons by their atmospheric magnetic field.  We also establish that for stars with an averaged magnetic field above $30$ kG for some axion models, this can lead to the production of an x-ray excess in their electromagnetic spectra. These stars will have an excess of x-ray electromagnetic radiation above $ 10^{-5}\; L_\odot$, if the axions have an axion-photon coupling constant above $5\; 10^{-11}{\rm GeV }^{-1}$ and an axion mass of the order of $10^{-7}{\rm eV}$. These values correspond to a region of the axion parameter space $m_a$ -- $g_{a\gamma}$ that is not been probed by current axion detectors including the CAST collaboration as reported in \citet[][]{2017NatPh..13..584A}. We predict that main sequence and post-main sequence stars can have x-ray luminosities due to axion emission above $10^{-5}\;L_\odot$ which is larger or comparable to the x-ray luminosity measured in some low-mass  stars \citep{ 2016MNRAS.462.4442S}.

 \medskip\noindent
Therefore, this work shows that if high precision stellar x-ray observations are made available by the next generation of satellites, it will be possible to constrain  the  $m_a$ -- $g_{a\gamma}$  parameters using another axion mechanism operating  low-mass stars, rather than the classical axion energy-loss that is known to affect the evolution of the  Sun and horizontal branch stars. 
By taking advantage of a large number of stars expected to be observed by the next generation of x-ray missions, this novel method should allow us to enhance the current axion constraints by using measurements of stellar x-ray luminosities. Moreover,  by using a large ensemble of stars including  more massive stars such result can possibly be extended for  axion models with masses varying from  $10^{-7}$ to $10^{-5}{\rm eV}$, a  region of the axion parameter space not continuously probed by current detectors. In that way, such study could complement the few but important measurements in that  axion mass range made by haloscopes.  

\medskip\noindent
Moreover, this study could also contribute and complement other methods to put constraints using the x-ray emission resulting from the interaction of stellar axions with the galactic magnetic field between the source of axion and the Earth. \citet{2020arXiv200803305D} using x-ray data from the Quintuplet and Westerlund 1 super star clusters, were able to found upper limits to the axion-photon coupling and axion mass with a $95\%$ confidence level:  $g_{a\gamma} \lesssim  3.6\times 10^{-12}\;{\rm GeV}^{-1}$ and $m_a \lesssim 5\times 10^{-11}\;{\rm GeV}$.
This constraint is obtained under the usual assumption  that the magnetic field is homogeneous between the location of stellar clusters and Earth.

\medskip\noindent
The existence of axions is complicated to establish mainly since the axion is a low mass particle with a weak interaction with standard ones. For this reason, many classic phenomena can mask the astrophysical signature of the axion. As in many other situations in physics, a possible discovery of axions will only be possible if we can find different astrophysical scenarios that can be explained by the same axion particle. Therefore the study of the x-ray spectra produced by the interaction of the axion with the magnetic fields, either in the atmosphere of the stars or in the galaxy, can contribute significantly to resolve this problem.

\medskip\noindent
Finally, this study is of interest for the future search of axion signatures in stellar astrophysics scenarios, namely when looking for signatures of the production of axions in stars. It is also useful to take advantage of a large amount of high-quality observational data that will be made available by some present, and future astronomical observatories and astrophysical space missions. 
As shown in this work, a prominent axion signature to look for is an excess of radiation in the x-ray energy band of the electromagnetic spectrum (that cannot be explained by the standard magnetism of stars) -- the next generation of x-ray satellites will be a powerful tool to test such class of axion models.  
Among others, we mention the Nuclear Spectroscopic Telescope Array mission or NuSTAR \citep[this  is  a x-ray telescope launched in 2012, operating  in the band from 3 to 79 keV][]{2013ApJ...770..103H}, and the Time-Resolving Observatory for Broadband Energy x-rays \citep[STROBE-X;][]{2019arXiv190303035R} mission. The latter is a future x-ray satellite expected to observe in the energy range: 2 to 60 KeV,   exactly the energy range where  axion models are expected to have the maximum emission.

\medskip\noindent
In this preliminary study, like in others found in literature, we do not consider the diversity of spectral properties of the stellar atmosphere  
and magnetic fields found in such low-mass  stars (and many others across the HR diagram) in our calculations. We also neglect the time variability observed in the magnetic fields of the Sun and many other stars. However, we expect  that a detailed prediction of the x-ray radiation emitted resulting from axion interaction with the stellar magnetic field should take such effects into account. Nevertheless, we  believe that the overall result found in this study should remain valid.

\medskip\noindent
\begin{acknowledgments}
The author is grateful to the  MESA  team for having made their code publicly available. The author also thanks the anonymous referee for the revision of the manuscript. I.L. thanks the Funda\c c\~ao para a Ci\^encia e Tecnologia (FCT), Portugal, for the financial support to the Center for Astrophysics and Gravitation (CENTRA/IST/ULisboa) 
through the Grant Project~No.~UIDB/00099/2020  and Grant No. PTDC/FIS-AST/28920/2017.
\end{acknowledgments}



\end{document}